\documentclass[pra,aps,twocolumn,showpacs,10pt]{revtex4-1}
\usepackage{amsmath,amssymb,graphicx,amsthm,dsfont,bm,color,ulem}

\begin{document}

\title{Couplings between the temporal and orbital angular momentum degrees of freedom in ultrafast optical vortices}

\author{Miguel A. Porras}

\affiliation{Grupo de Sistemas Complejos, ETSIME, Universidad Polit\'ecnica de Madrid, Rios Rosas 21, 28003 Madrid, Spain}

\author{Claudio Conti}

\affiliation{Institute for Complex Systems (ISC-CNR), Via dei Taurini 19, 00185 Rome, Italy\\
Department of Physics, University Sapienza, Piazzale Aldo Moro 5, 00185 Rome, Italy}

\begin{abstract}
  In any form of wave propagation, strong spatiotemporal coupling appears when non-elementary, three-dimensional wave-packets are composed by superimposing pure plane waves, or spontaneously generated by light-matter interaction and nonlinear processes. Ultrashort pulses with orbital angular momentum (OAM), or ultrashort vortices, furnish a critical paradigm in which the analysis of the spatiotemporal coupling in the form of temporal-OAM coupling can be carried out accurately by analytical tools.
  By generalizing and unifying previously reported results, we show that universal and spatially heterogeneous space-time correlations occur in propagation-invariant temporal pulses carrying OAM.
  In regions with high intensity, the pulse duration has a lower bound fixed by the topological charge of the vortex and such that the duration must increase with the topological charge. In regions with low intensity in the vicinity of the vortex, a large blue-shift of the carrier oscillations and an increase of the number of them is predicted for strongly twisted beams. We think that these very general findings highlight the existence of a structural coupling between space and time. These results have also applications
  in free-space communications, spectroscopy, and high-harmonic generation.
\end{abstract}

\maketitle

\section{Introduction}

Since their first introduction in acoustics \cite{LU1992}, propagation-invariant, three-dimensional wave-packets attracted a lot of attention because of fundamental issues, as superluminality \cite{SAARI1997,RECAMI2008} and wave-function localization \cite{BIRULA1996,SAARI2006,SAARI2010}, and potential applications as in telecommunications \cite{LU1999, KONDAKCI2019}. Solutions of wave-equations evolving without distortion were also considered in Bose-Einstein condensation \cite{CONTI2008}, linear and non linear optics \cite{PORRAS2003, CONTI2003}, and more recently in polaritonics \cite{SANVITTO2018, COLAS2019} and  hydrodynamics \cite{STEER2019}.
X waves with orbital angular momentum (OAM) have been recently described as solutions of Maxwell equations \cite{CONTI1,CONTI2}, and quantum optical X waves \cite{CIATTONI2007} with OAM have been studied for their applications in free-space multilevel transmission systems (see \cite{ORNIGOTTI2017} and references therein).

In spite of the relevance of X waves with OAM as the first reported ultrashort vortices with diffraction-free behavior, actual linear and nonlinear experiments with vortices in ultrafast (femtosecond or attosecond) regimes employ pulsed Laguerre-Gauss (LG) beams \cite{REGO2019,GARIEPY2019,HERNANDEZ,GENEAUX,REGO}. There is increasing interest in synthesizing shorter and shorter vortices of the LG type, \cite{BEZUHANOV,ZEYLIKOVICH,TOKIZANE,SHVEDOV,RICHTER,YAMANE} which in the high intensity regime are used to generate high-harmonics and attosecond vortices with large OAM \cite{REGO2019,GARIEPY2019,HERNANDEZ,GENEAUX,REGO}. From a theoretical point of view, it has been recently demonstrated \cite{PORRAS1,PORRAS2,PORRAS2019} that pulsed LG beams with the so-called isodiffracting conditions \cite{PORRAS6} maintain a propagation-invariant pulse temporal shape as X-waves, even if they are subject to diffraction.

Notwithstanding these many investigations, fundamental questions regarding ultrafast X and LG vortices remain unsolved. In recent years, a coupling between the temporal and OAM degrees of freedom in ultrashort (few-cycle and sub-cycle) pulses has been described theoretically in ultrafast X vortices \cite{CONTI1,CONTI2} and LG vortices \cite{PORRAS1,PORRAS2,PORRAS2019}. A consequence detailed in these works is that an arbitrarily short pulse cannot carry an arbitrarily high OAM, but there is a lower bound to its duration for given OAM. However, the effects of temporal-OAM coupling in ultrashort X and LG vortices appear to be quite different, even contradictory. The results in \cite{CONTI1} show that in ultrafast X vortices the carrier frequency and the number of oscillations increase with angular momentum, and that the increase of the carrier frequency is faster than the increase of the number of oscillations so that the pulse duration decreases with OAM. On the contrary, \cite{PORRAS1,PORRAS2} report an absent or negligible blue shift for ultrafast LG vortices and an increase of the pulse duration with OAM.

Here we report a general and unified treatment for both types of ultrashort vortices with propagation-invariant pulse shape. First we show that the two analyses are not in contradiction because they refer to different spatial regions of the ultrashort vortex. Second we unveil a very rich phenomenology common to all ultrafast vortices with propagation-invariant pulse shape. We distinguish the effects of the temporal-OAM coupling depending on the model from those intrinsic or ``universal".
Our results have implications in perturbative nonlinear optics and high-field, nonperturbative, light-matter interactions (as in high-harmonic and attosecond pulse generation experiments) and in quantum optics in the low photon regime. Indeed, in ultrafast propagation of twisted pulses, regions of high intensity (high energetic ring) and low intensity (central vortex core with phase singularity) are present, and the nature of temporal-OAM coupling varies a lot in these different portions of these three-dimensional wavepackets. We describe below different experimentally testable features to demonstrate the heterogeneous spatiotemporal coupling in ultrashort vortices.

\section{Ultrafast Laguerre-Gauss and X vortices}

\begin{figure}[!]
  \centering
  \includegraphics[width=7cm]{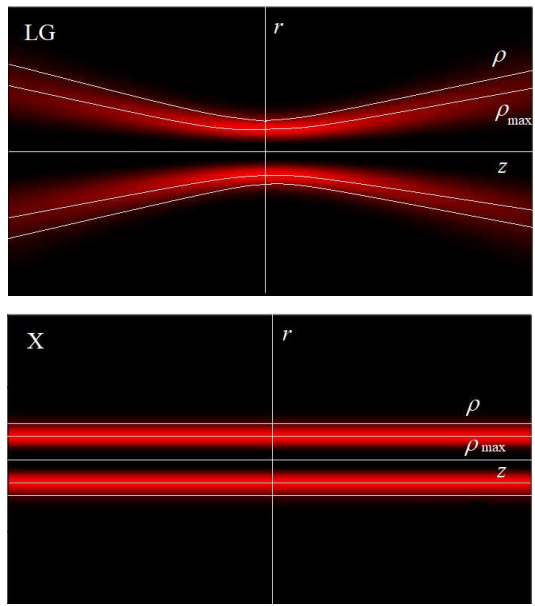}
  \caption{The white curves represent sagittal sections of the revolution hyperboloids $\rho=\mbox{const.}$ of ultrafast LG vortices (top) and of the cylinders $\rho=\mbox{const.}$ of ultrafast X vortices, called here caustic surfaces, along which the pulse shape is invariant. The red colors illustrate the geometry of the fluence distribution of both type of vortices.}\label{Fig1}
\end{figure}

We express the optical field $E(r,\phi,t,z)$, or $E$ for short, of an ultrafast or ultrashort vortex of topological charge $l$, or $l$ units of OAM, as the superposition of monochromatic vortex with charge $l$ and varying frequency. The analytical signal complex representation \cite{BORNWOLF} is
\begin{equation}\label{ERHOT}
E=E(\rho,t')e^{il\phi} = \frac{1}{\pi} \int_0^\infty d\omega \hat E_\omega(\rho) e^{-i\omega t'} e^{il\phi}\,.
\end{equation}
For ultrafast LG vortices of charge $l$ and zero radial order,
\begin{equation}\label{LG}
\hat E_\omega(\rho) = \hat a_\omega \hat E_\omega^{\rm LG}(\rho)=\hat a_\omega D\left(2\omega\rho^2\right)^{\frac{|l|}{2}}e^{-\omega\rho^2}\,,
\end{equation}
where $\hat a_\omega$ specifies the weights or spectrum of monochromatic LG beams, $D = e^{-i(|l|+1)\tan^{-1}(z/z_R)}/\sqrt{1+(z/z_R)^2}$ accounts for Gouy's phase shift and attenuation due to diffraction spreading, $z_R$ is the Rayleigh distance, $c$ is the speed of light in vacuum, $\rho=r/\sqrt{2z_Rc[1+(z/z_R)^2]}$ is a normalized radial coordinate at each propagation distance $z$, $(r,\phi,z)$ are cylindrical coordinates, and $t'=t-z/c-\rho^2z/z_R$ is a local time that is equal to zero at the instant of arrival of the pulse at any position $(\rho,z)$ \cite{PORRAS1,PORRAS2}. A constant value of $\rho$ represents the revolution hyperboloid, or caustic surface, $r=\rho \sqrt{2z_Rc[1+(z/z_R)^2]}$ whose revolution axis is the $z$ axis, as illustrated in the top panel of Fig. \ref{Fig1}. According to Eqs. (\ref{ERHOT}) and (\ref{LG}), the pulse shape $E(\rho,t')$ only depends on $\rho$ (except for the global complex amplitude factor $D$), and therefore does not change with propagation distance $z$ along a given revolution hyperboloid.

For X waves, or ultrashort X vortices of topological charge $l$, the optical field $E(r,\phi,t,z)$ is given by Eq. (\ref{ERHOT}) with
\begin{equation}\label{X}
\hat E_\omega(\rho) = \hat a_\omega \hat E_\omega^{\rm X}(\rho) = \hat a_\omega J_l(\omega\rho)\,,
\end{equation}
where $J_l(x)$ is the Bessel function of the first kind and order $l$, $\hat a_\omega$ is the spectrum of monochromatic Bessel beams, $\rho=(\sin\theta/c)r$, $t'=t-(\cos\theta/c)z$ is the local time for the superluminal propagation at speed $c/\cos\theta$, and $\theta$ is the cone angle. The pulse shape changes with the radius $\rho$, but since X waves are diffraction-free, the pulse shape does not change along the cylinders $\rho=\mbox{const.}$ coaxial with the $z$ axis drawn in the bottom panel of Fig. \ref{Fig1}.

For completeness and future generalizations to propagation in linear and nonlinear media, the optical field $E(r,\phi,t,z)$ of ultrashort X vortices satisfies the exact wave equation in free space, $\Delta E- (1/c^2)\partial^2_t E=0$, which written in terms of the local time reads $\Delta E =(2/c)\partial^2_{z,t'} E$. The optical field of ultrashort LG vortices satisfies the paraxial version of the last equation, also called pulsed beam equation or paraxial wave equation for ultrashort wave forms \cite{HEYMAN,PORRASJ,BESIERIS}, obtained by neglecting the second derivative with respect to $z$ in the Laplace operator, i. e., $\Delta_\perp E =(2/c)\partial^2_{z,t'} E$, where $\Delta_\perp = \partial^2_x+ \partial^2_y$. The temporal-OAM coupling is a particular spatiotemporal coupling involving the azimuthal and temporal degrees of freedom of the wave. Coupling means here the azimuthal and temporal structure of the wave are not independent as imposed by the wave equation.  Spatiotemporal couplings in the solutions of the wave equation are well-known from decades. They are usually small for few-cycle pulses without OAM, e. g., for the fundamental pulsed Gaussian beams \cite{PORRAS6}, although can be artificially enhanced for practical purposes \cite{AKTURK}, and appear to be much more pronounced with pulsed carrying OAM, as described in \cite{CONTI1,PORRAS1}.

Ultrafast LG and X vortices with $z_R$ and $\theta$ independent of frequency are the only ultrafast vortices that have been described whose temporal shape does not change on propagation (although it changes from one radius to another). In these two types of vortices with propagation-invariant temporal shape {\it seemingly discordant} temporal-OAM couplings have been described. The different definitions and dimensions of the normalized radial distance with the same symbol, $\rho$, for these two types of ultrashort vortices will help us to visualize more clearly that the temporal-OAM couplings are substantially the same. The limitation to $z_R$ and $\theta$ independent of frequency allows to distinguish the intrinsic and unavoidable effects of OAM on pulse temporal shape from those arising from propagation in more general models. Mixed propagation-OAM effects on pulse temporal shape in these more general models will be investigated once pure effects of OAM are fully understood.

In the following we will characterize any of the above functions of frequency, say $\hat f_\omega$, by its mean frequency
\begin{equation}\label{MEAN}
\bar \omega_f = \frac{\int_0^{\infty}d\omega |\hat f_\omega|^2 \omega}{\int_0^{\infty}d\omega |\hat f_\omega|^2} \,,
\end{equation}
its Gaussian-equivalent half bandwidth
\begin{equation}\label{GEBW}
\Delta \omega_f = 2\left[\frac{\int_0^{\infty}d\omega |\hat f_\omega|^2 (\omega-\bar\omega_f)^2}{\int_0^{\infty}d\omega |\hat f_\omega|^2} \right]^{1/2} \,,
\end{equation}
(yielding the $1/e^2$ decay of $|\hat f_\omega|^2$ for Gaussian-like $\hat f_\omega$), and the corresponding pulse shape in time domain,
\begin{equation}
f(t')=\frac{1}{\pi}\int_0^\infty d\omega \hat f_\omega e^{-i\omega t'}
\end{equation}
by its mean time $\bar t'_f$ and half duration $\Delta t_f$, defined in the same way for $|f(t')|^2$. The term ``pulse shape" is often synonymous of pulse (complex) envelope, but our analysis applies also to pulses of duration below the single-cycle regime, as defined in \cite{BRABEC}, for which the concept of envelope is physically meaningless \cite{BRABEC}. Thus, pulse shape will refer to the instantaneous time-evolution of the optical signal $f(t')$. For functions that depend also on $\rho$, we will specify the radius where the above quantities are evaluated, e. g., $\Delta \omega_E(\rho)$ for the bandwidth of $\hat E_\omega(\rho)$ at $\rho$, or $\Delta t_E(\rho)$ for the duration of $E(\rho,t')$ at $\rho$.

\section{Temporal-OAM couplings at the bright caustic surface}

In experiments involving nonlinear interactions, the region in the transversal plane where the pulse energy is high is the most relevant. This is why \cite{PORRAS1,PORRAS2} considered the energy density
\begin{equation}\label{E}
{\cal E}(\rho) = \int_{-\infty}^\infty \!\!\!\!dt' (\mbox{Re}E)^2 = \frac{1}{2}\int_{-\infty}^\infty \!\!\!\!dt'|E|^2 = \frac{1}{\pi}\int_0^\infty \!\!\!\!d\omega |\hat E_\omega(\rho)|^2 \,,
\end{equation}
{whose typical spatial distribution is shown in Fig. \ref{Fig1}, and analyze the pulse properties at the hyperbolic or cylindrical caustic surface, $\rho_{\rm max}$, where the energy density is maximum, or bright caustic surface. As seen in the examples of Figs. \ref{Fig2}(a) and \ref{Fig3}(a), $\rho_{\rm max}$ increases monotonically with $|l|$.

\subsection{Ultrafast LG vortices}

A fundamental restriction to the pulse characteristics at the bright caustic surface of all ultrafast LG vortex has been recently described \cite{PORRAS2}. With given topological charge $l$, the relative spectral bandwidth of the pulse at the bright caustic surface necessarily satisfies inequality $\Delta\omega_E(\rho_{\rm max})/\bar\omega_E(\rho_{\rm max})\leq 2/\sqrt{|l|}$. More recently, the mean frequency of the oscillations at this caustic, $\bar\omega_E(\rho_{\rm max})$, has been shown to coincide with the mean frequency $\bar \omega_a$ of the spectrum of LG beams $\hat a_\omega$ in the particular situation that $\hat a_\omega$ is chosen to be a power-exponential spectrum \cite{PORRAS1}, but, as argued below, the approximate equality $\bar\omega_E(\rho_{\rm max})\simeq \bar \omega_a$ holds with generality. Thus, in practice, all ultrafast LG vortices satisfy
\begin{equation}\label{UB}
\frac{\Delta\omega_E(\rho_{\rm max})}{\bar\omega_a}\lesssim \frac{2}{\sqrt{|l|}}\,.
\end{equation}
Since $\Delta t_f\Delta\omega_f\gtrsim 2$ for any pair $\hat f_\omega$ and $f(t')$, the approximate equality being only reached for Gaussian-like $f_\omega$, it follows that the duration of the pulse at the bright caustc is restricted by $\Delta t_E(\rho_{\rm max})\gtrsim \sqrt{|l|}/\bar\omega_E(\rho_{\rm max})$. On account that $\bar\omega_E(\rho_{\rm max})\simeq \bar \omega_a$, the lower bound to the pulse duration at the bright caustic is in practice
\begin{equation}\label{LB}
\Delta t_E(\rho_{\rm max})\gtrsim \frac{\sqrt{|l|}}{\bar\omega_a}\,.
\end{equation}
For a particular application, one may wish to fix a pulse shape of certain characteristics at the bright caustic. The above restrictions then impose that the topological charge of an ultrafast vortex with such a pulse shape has the upper bound $|l|\lesssim [\bar\omega_E^2(\rho_{\rm max}) \Delta t_E^2(\rho_{\rm max})]$, where the square bracket means integer part. A consequence of the dependence on $l$ of the lower bound to the pulse duration in inequality (\ref{LB}) is that the pulse at the bright caustic of the ultrafast LG vortex synthesized with the same spectrum $\hat a_\omega$ necessarily broadens with increasing magnitude of the topological charge, a temporal-OAM coupling effect has been recently verified in \cite{PORRAS1} with a particular choice of $\hat a_\omega$. These bounds settles unavoidable limits, for example, to the velocity of information transmission or to the number of channels in communication systems based on ultrashort pulses carrying OAM.

\begin{figure}[!]
  \centering
  \includegraphics[width=8.5cm]{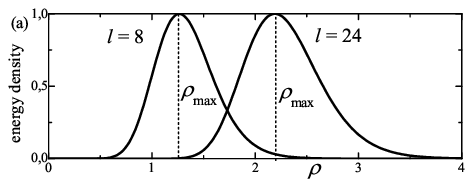}
  \includegraphics[width=4.2cm]{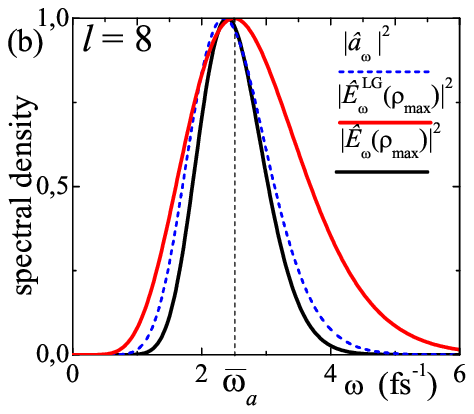} \includegraphics[width=4.2cm]{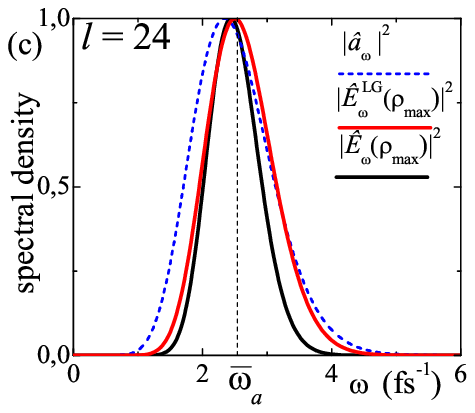}
  \caption{Ultrashort LG vortices at their bright caustic surface. In these two examples $\hat a_\omega=\omega^{7.5} e^{-3.2\omega}$, yielding an about single-cycle pulse in time domain of mean frequency $\bar\omega_a=2.5$ fs$^{-1}$ in the near infrared. (a) Energy density radial profiles for $l=8$ and $l=24$. The vertical lines are located at $|l^2|/2\omega_a$, which accurately locate $\rho_{\rm max}$. (b) and (c) Factors $|\hat a_\omega|^2$ (dashed blue curve) and $|\hat E_\omega^{\rm LG}(\rho_{\rm max})|^2$ (red curve) of the spectral density $|\hat E_\omega(\rho_{\rm max})|^2$ (black curve) at the bright caustic. Since absolute values are irrelevant, all functions are normalized to their peak values. }\label{Fig2}
\end{figure}

A intuitive explanation of these temporal-OAM coupling effects for ultrashort LG vortices with general $\hat a_\omega$ (and that applies also to the ultrashort X vortices considered below) is as follows. The spectral density of the ultrafast LG vortex in (\ref{LG}) is the product of $|\hat a_\omega|^2$, centered about $\bar\omega_a$, and $|\hat E_\omega^{\rm LG}(\rho)|^2$, characterized by a single maximum at $\omega_M= |l|/2\rho^2$. At arbitrary $\rho$ these two functions of frequency do not overlap. The energy density in Eq. (\ref{E}) is maximum at $\rho_{\rm max}$ because these two functions overlap optimally at this radius, as in the examples of Figs. \ref{Fig2}(b) and (c), which implies that the mean frequency of the ultrafast LG vortex at the bright caustic, $\bar\omega_E(\rho_{\rm max})$, is approximately equal to $\omega_M$ and to $\bar\omega_a$, i. e., no significant blue of red shift of the mean frequency at the bright caustic is expected, as described in \cite{PORRAS2} for a specific $\hat a_\omega$. Relation $\bar\omega_a\simeq\omega_M =|l|/2\rho_{\rm max}^2$ provides the approximate expression $\rho_{\rm max}^2\simeq |l|/2\bar\omega_a$ for  the location of the bright, as verified in the examples of Fig. \ref{Fig2}(a). At this radius $|\hat E_\omega^{\rm LG}(\rho_{\max})|^2= |l|^{|l|}(\omega/\bar\omega_a)^{|l|}e^{-|l|\omega/\bar\omega_a}$ is an approximate Gaussian function, specially for large $|l|$, of Gaussian width $\Delta\omega_{\rm LG}(\rho_{\rm max}) \simeq 2\bar\omega_a/\sqrt{|l|}$. Thus, as the product of $|\hat a_\omega|^2$ and $|\hat E_\omega^{\rm LG}(\rho_{\max})|^2$, the spectral density $|\hat E_\omega(\rho_{\rm max})|^2$ cannot be wider than $|\hat E_\omega^{\rm LG}(\rho_{\max})|^2$, i. e., $\Delta\omega_{E}(\rho_{\rm max})\leq \Delta\omega_{\rm LG}(\rho_{\rm max})$, or $\Delta\omega_{E}(\rho_{\rm max})/\bar\omega_a\lesssim 2/\sqrt{|l|}$ and $\Delta t_E(\rho_{\rm max})\gtrsim \sqrt{|l|}/\bar\omega_a$, as in inequalities (\ref{UB}) and (\ref{LB}). Although approximate equalities are used in this derivation, the result is the same as the one rigorously derived in \cite{PORRAS1}. Also, with given $\hat a_\omega$ condition $\Delta t_E(\rho_{\rm max})\gtrsim \sqrt{|l|}/\bar\omega_a$ implies that the pulse duration at the bright caustic necessarily increases from the lowest value $\Delta t_a$ (for $l=0$) as $|l|$ increases, as described in \cite{PORRAS2} but for general $\hat a_\omega$.

\subsection{Ultrafast X vortices}

Similar reasonings applied to ultrafast X vortices, or X wave with OAM, allow us to conclude that temporal-OAM couplings at theircylindrical caustic surface of maximum energy density are qualitatively similar. In particular, there is a lower bound to their duration.

\begin{figure}[t!]
  \centering
  \includegraphics[width=8.5cm]{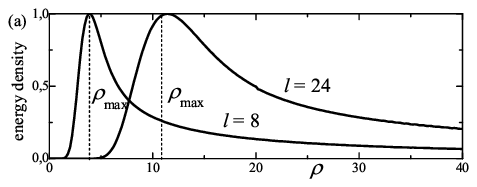}
  \includegraphics[width=4.2cm]{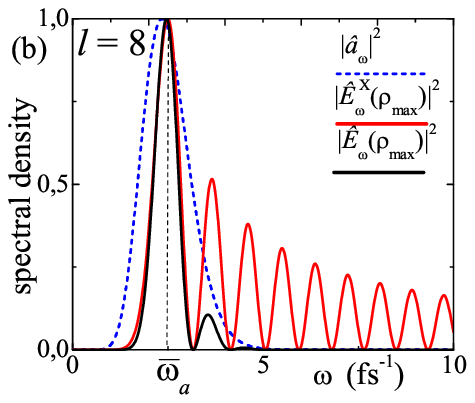} \includegraphics[width=4.2cm]{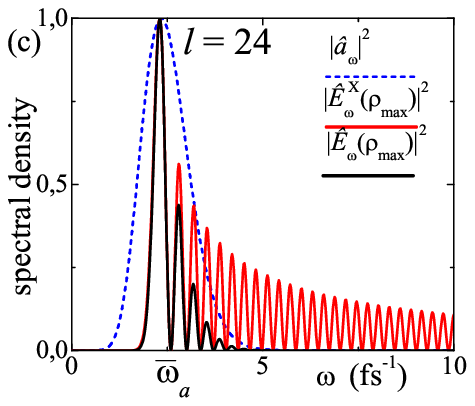}
  \includegraphics[width=4.2cm]{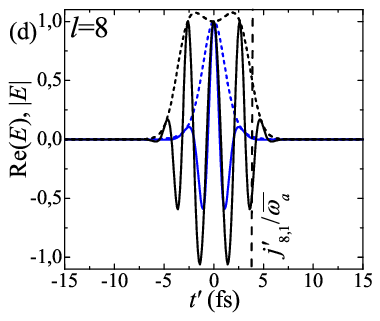} \includegraphics[width=4.2cm]{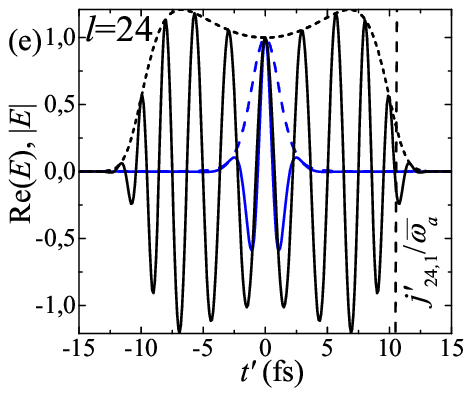}
  \caption{Ultrasfast X vortices at their bright caustic surface. In these two examples $\hat a_\omega=\omega^{7.5} e^{-3.2\omega}$, as for the ultrafast LG vortices of Fig. \ref{Fig2} (about single-cycle pulse of mean frequency $\bar\omega_a=2.5$ fs$^{-1}$) (a) Energy density radial profiles for $l=8$ and $l=24$. The vertical lines are located at $j'_{l,1}/\bar\omega_a$, and approximate $\rho_{\rm max}$. (b) and (c) Factors $|\hat a_\omega|^2$ (dashed blue curve) and $|\hat E_\omega^{\rm X}(\rho_{\rm max})|^2$ (red curve) of the spectral density $|\hat E_\omega(\rho_{\rm max})|^2$ (black curve) at the bright caustic. Since absolute values are irrelevant, all functions are normalized to their peak values.
  (d) and (e) Numerically evaluated real field (solid curves) and envelope (dashed curves) of the pulse with the spectrum $\hat a_\omega$ (blue curves), and of the X wave with the same spectrum and $l=8$ and with $l=24$ at their bright caustic (black curves). All functions are normalized to the value at $t'=0$.
   }\label{Fig3}
\end{figure}

\begin{figure}[b]
  \centering
  \includegraphics[width=4.2cm]{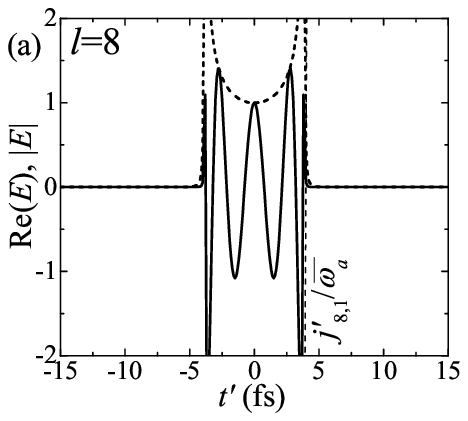} \includegraphics[width=4.2cm]{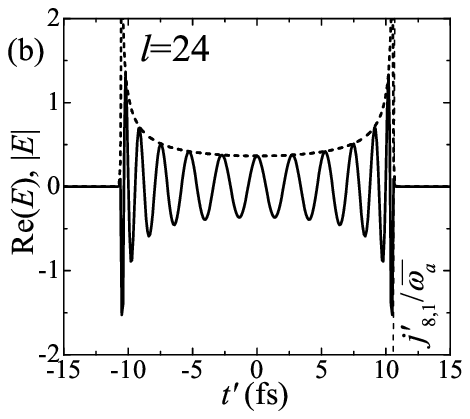}
  \caption{Real field (solid) and modulus (dashed) of minimal pulse shapes at the bright caustic of ultrafast X vortices as given by Eq. (\ref{FJL}) with $\rho_{\rm max}=j'_{l,1}/\bar\omega_a$ and $\bar\omega_a=2.5$ fs$^{-1}$. Their half duration is just $\rho_{\rm max}$.}\label{Fig4}
\end{figure}

\begin{figure}[t!]
  \centering
  \includegraphics[width=4.2cm]{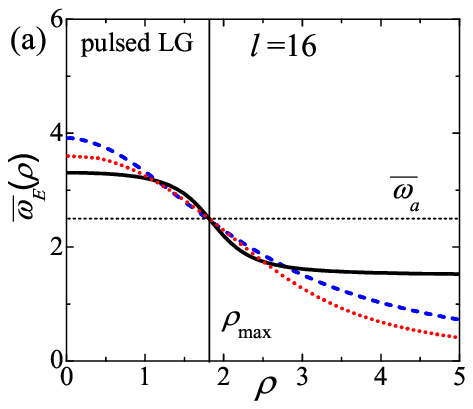} \includegraphics[width=4.2cm]{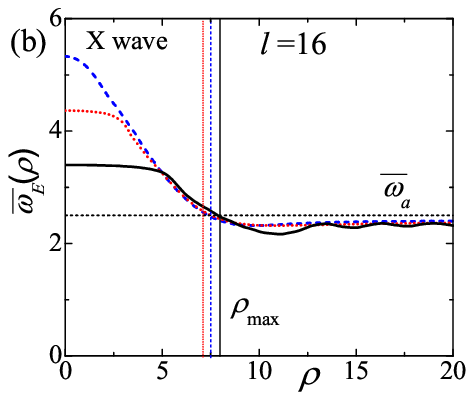}
  \includegraphics[width=4.2cm]{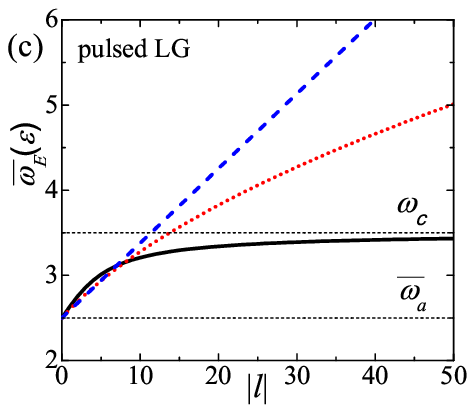} \includegraphics[width=4.2cm]{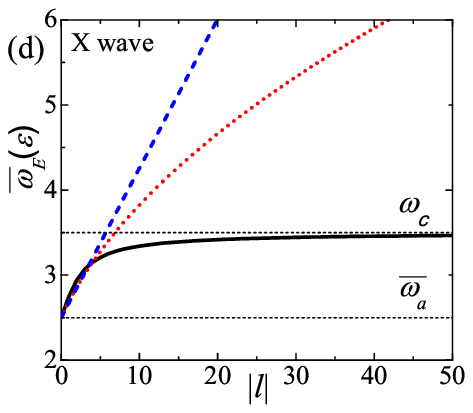}
  \caption{Behavior close to the vortex: Mean frequencies of ultrashort LG and X vortices with three different spectra $\hat a_\omega$: Power-exponential $\hat a_\omega \propto\omega^{\alpha-1/2}\exp(-\alpha \omega/\bar\omega_a)$ with $\alpha=14.24$ (dashed blue curves), Gaussian $\hat a_\omega \propto \exp[-(\omega-\bar\omega_a)^2/\Delta\omega_a^2]$ with $\Delta\omega_a\simeq 1$ fs$^{-1}$ (red dotted curves), and square $\hat a_\omega \propto 1$ if $|\omega-\bar\omega_a|<\delta\omega$ and $0$ otherwise, with $\delta\omega \simeq 1$ fs$^{-1}$ (black solid curves), all them with the same mean frequency $\bar\omega_a=2.5$ fs$^{-1}$. The three spectra correspond to single-cycle pulses $a(t')$ (one oscillation in the full-width at half-maximum of the cycle-averaged intensity $|a(t')|^2$) of different shapes. (a) and (b) Mean frequency as a function of radial distance for ultrashort LG and X vortices with $l=16$. The radius $\rho_{\rm max}$ is almost independent of these three spectral shapes for LG vortices, and appreciably varies with spectral shape for X vortices (vertical lines). (c) and (d) Mean frequency in the vicinity of the vortex, $\rho=\epsilon$, for ultrafast LG and X vortices with the above spectral shapes $\hat a_\omega$ as functions of the magnitude of the topological charge $|l|$.} \label{Fig5}
\end{figure}

Figure \ref{Fig3}(a) shows two radial profiles of the energy density featuring single maxima at certain radii $\rho_{\rm max}$ that increase with $|l|$. The spectral density of the ultrashort X vortex is also the product of $|\hat a_\omega|^2$ and $|\hat E_\omega^{\rm X}(\rho)|^2 = J_l^2(\omega\rho)$, the latter characterized by an absolute maximum at $\omega_M=j'_{l,1}/\rho$, where $j'_{l,1}\simeq |l|+ 0.80869 |l|^{1/3}$ is the location of the first maximum of $J_l(x)$ \cite{ABRAMOVITZ}. As illustrated in Fig. \ref{Fig3}(b) and (c), optimum overlapping of these two functions at the radius $\rho_{\rm max}$ of maximum energy density implies again that $\bar\omega_{E}(\rho_{\rm max})\simeq \bar\omega_X(\rho_{\rm max})\simeq \bar\omega_a$. We then do not expect significant red or blue shift of the mean frequency at the bright caustic with respect to the value $\bar\omega_a$ determined by the spectrum of Bessel beams, as for ultrashort LG vortices \cite{PORRAS2},
and at variance with ultrashort X vortices in the neighborhood of their vortex \cite{CONTI1}. Relation $\rho_{\rm max}\simeq j'_{l,1}/\bar\omega_a$ provides an approximation to the location of the bright caustic, as is verified in Fig. \ref{Fig3}(a). As the product of $\hat a_\omega$ and $|\hat E_\omega^X(\rho_{\rm max})|^2=J_l^2(\omega j'_{l,1}/\bar \omega_a)$, the full spectral density $|\hat E_\omega(\rho_{\rm max})|^2$ at the bright caustic cannot be wider than $J_l^2(\omega j'_{l,1}/\bar \omega_a)$, approaching it only with wider and wider $|\hat a_\omega|^2$, in which case the narrowest pulse at the bright caustic is $E(\rho_{\rm max},t')= (1/\pi)\int_0^{\infty} d\omega J_l(\omega\rho_{\rm max})e^{-i\omega t'}$ with $\rho_{\rm max}\simeq j'_{l,1}/\bar\omega_a$, i. e.,
\begin{equation}\label{FJL}
E(\rho_{\rm max},t')= \lim_{\epsilon\rightarrow 0}\displaystyle\frac{\left[\sqrt{(\epsilon+it')^2+\rho_{\rm max}^2}-(\epsilon+it')\right]^{|l|}}{\pi\rho_{\rm max}^{|l|}\sqrt{(\epsilon+it')^2+\rho_{\rm max}^2}}\, ,
\end{equation}
(for $l>0$, and multiplied by $(-1)^{|l|}$ for $l<0$) with $\rho_{\rm max}\simeq j'_{l,1}/\bar\omega_a$. As clearly seen in Fig. \ref{Fig4}, the half duration of the pulses in Eq. (\ref{FJL}) is $\Delta t=\rho_{\rm max}$; hence the minimum pulse duration at the bright caustic of an ultrashort X vortex with $l$ units of OAM is $\rho_{\rm max} \simeq j'_{l,1}/\bar \omega_a$, and in conclusion, for any other ultrashort X vortex
\begin{equation}\label{LX}
\Delta t_E(\rho_{\rm max}) \gtrsim \frac{j'_{l,1}}{\bar\omega_a}\simeq \frac{|l|+ 0.80869 |l|^{1/3}}{\bar\omega_a} \simeq  \frac{|l|}{\bar\omega_a}\,,
\end{equation}
the last approximate equality being valid for large $|l|$. Also, the $l$-dependence of this lower bound implies that with given $\hat a_\omega$ the pulse shape at the bright caustic must necessarily increase as $|l|$ grows; indeed it must do almost linearly above a certain value of $|l|$. This description of the temporal-OAM coupling effects at the bright caustic of ultrashort X vortices complements the temporal-OAM coupling effects close to the vortex described in \cite{CONTI1}, and its validity is tested in the examples of Figs. \ref{Fig3}(d) and (e). With the same spectrum $\hat a_\omega$ of about a single-cycle pulse (blue curves), the mean frequency at the bright caustic does not significantly depart from $\bar\omega_a$ as $|l|$ grows (black curves), but the number of oscillations increases with $|l|$ so as the pulse duration is always above  $j'_{l,1}/\bar \omega_a$ (vertical dashed lines). Broadening is very small for durations $\Delta t_a$ of $a(t')$ well-above the lower bound, i. e., for $\Delta t_a \gg j'_{l,1}/\bar \omega_a$, but is very pronounced for $\Delta t_a \lesssim j'_{l,1}/\bar \omega_a$, as in Figs. \ref{Fig3}(d) and (e), in which case the pulse shape is an apodized, non-singular version of the minimal pulse shape in Fig. \ref{Fig4}.

As a partial conclusion, the temporal-OAM coupling effects at the more energetic caustic in the only two known cases of ultrafast vortices with propagation-invariant pulse shape are qualitatively similar. In particular, the mean or carrier frequency of $\hat a_\omega$ (which can be identified with the frequency of the laser source) is also the observable carrier frequency at the bright caustic of the produced ultrafast LG or X vortex. For ultrashort X vortices there also exists a lower bound to their duration at their bright caustic. Temporal broadening with increasing magnitude of the topological charge is more pronounced for X vortices than for LG vortices because of respective square root [Eq.~(\ref{LB})] and almost linear dependence [Eq.~(\ref{LX})] of the lower bound to the temporal duration.

\begin{figure}[!]
  \centering
  \includegraphics[width=4.2cm]{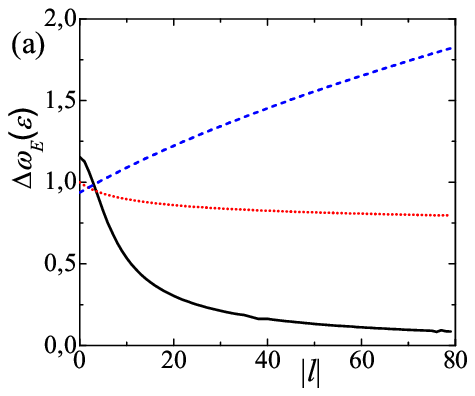} \includegraphics[width=4.2cm]{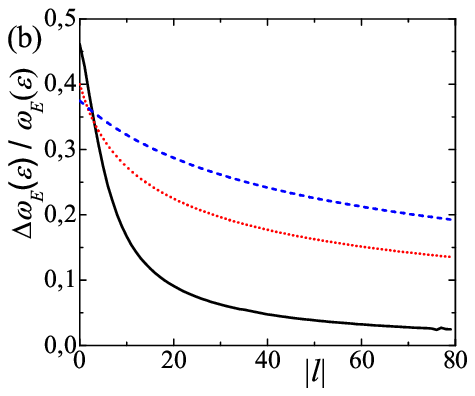}
  \includegraphics[width=8.4cm]{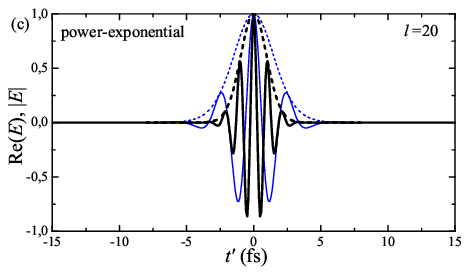} \\ \includegraphics[width=8.4cm]{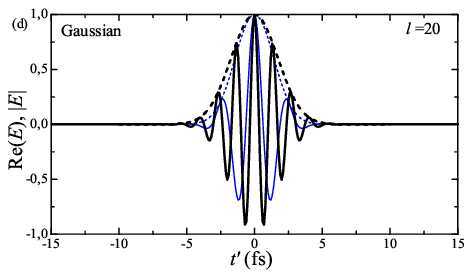}\\ \includegraphics[width=8.4cm]{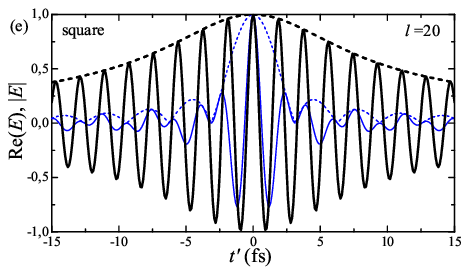}
  \caption{Behavior close to the vortex: Bandwidths and relative bandwidths of ultrashort LG vortices with the same power-exponential (dashed blue), Gaussian (red dotted), and square (black solid curves) spectra $\hat a_\omega$ as in Fig. \ref{Fig5}, all them corresponding to single-cycle pulses $a(t')$ of the same mean frequency $\bar\omega_a=2.5$ fs$^{-1}$. (a) Bandwidth and (b) relative bandwidth as functions of the magnitude of the topological charge.  (c), (d) and (e) Pulse shapes and envelopes for $l=20$ close to the vortex in the above three cases (black curves), compared to the single-cycle pulses $a(t')$ (blue curves). Depending on the specific $a(t')$, the pulse may broaden or shrink with $|l|$, but the number of oscillations always grows with $|l|$.}
  \label{Fig6}
\end{figure}

\section{Temporal-OAM couplings in the neighborhood of the vortex}

The low intensity region close to the vortex is of interest in quantum optics. Spectral anomalies in the vicinity of vortices of continuous, coherent and incoherent light have been studied in \cite{KLIMOV2009}. Temporal-OAM coupling effects in the vicinity of the vortex of X waves have been analyzed in detail in \cite{CONTI1} and \cite{CONTI2}. In turns out from our analysis that these temporal-OAM coupling effects are qualitatively the same of ultrahsort LG vortices in the proper spatial regions.

In the vicinity of the vortex, monochromatic LG beams behave asymptotically as $\hat E_\omega^{\rm LG}(\rho)\simeq (\omega\rho^2)^{|l|/2}$, and monochromatic Bessel beams as $\hat E_\omega^{\rm X}(\rho)\simeq (\omega\rho)^{|l|}$ (aside a irrelevant constant). The complete spectra are then $\hat E_\omega(\rho)\simeq \hat a_\omega(\omega \rho^2)^{|l|/2}$ for ultrashort X vortices and $\hat E_\omega(\rho)\simeq \hat a_\omega(\omega \rho)^{|l|}$ for ultrashort X vortices.

For the analysis below, we define
\begin{equation}\label{WL}
\bar \omega(m) = \frac{\int_0^\infty d\omega |\hat a_\omega|^2 \omega^{m+1}}{\int_0^\infty d\omega |\hat a_\omega|^2 \omega^{m}}\,,
\end{equation}
for any natural number $m$, with the dimensions of a frequency. In particular $\bar\omega(0)=\bar\omega_a$. It is readily seen that the combination $\bar\omega(m)\left[\bar\omega(m+1)-\bar\omega(m)\right]$ is the variance of the distribution $|\hat a_\omega|^2\omega^m$, and as such is positive (except if $\hat a_\omega$ contains a single frequency). It then follows that $\bar\omega(m)$ is a strictly growing function of $m$ if $\hat a_\omega$ is not a monochromatic spectrum. For all bell-shaped spectral shapes $\hat a_\omega$ of interest, $\bar\omega(m)$ is a concave function of $m$, and therefore $\bar\omega(m+1)/\bar\omega(m)$ approaches unity from above as $m$ increases.

In \cite{CONTI1} a linear increase of the mean frequency about the vortex of ultrafast X vortices with a power-exponential spectrum $\hat a_\omega$ is predicted. Growth of the mean frequency is a general feature, also for ultrafast LG vortices. Figures \ref{Fig5}(a) and (b) illustrate the behavior of the mean frequency as a function of the radial distance $\rho$ from the vortex for ultrafast LG and X vortices with three different choices of $\hat a_\omega$, all three corresponding to single-cycle pulses of different shapes (power exponential, Gaussian and square spectrum in a certain bandwidth, see caption for details). The mean frequency is always seen to grow from $\bar \omega_a$ at the bright caustic up to a locally constant value in the vicinity of the vortex. Indeed, with the spectral densities $|\hat E_\omega(\rho)|^2\simeq |\hat a_\omega|^2(\omega \rho^2)^{|l|}$ and $|\hat E_\omega(\rho)|^2\simeq |\hat a_\omega|^2(\omega \rho)^{2|l|}$ of ultrafast LG and X vortices close to the vortex, the mean frequency in this region is found to be
\begin{equation}
\bar \omega_E(\epsilon) = \bar\omega(|l|)\,, \quad \bar \omega_E(\epsilon) = \bar\omega(2|l|)\,,
\end{equation}
($\epsilon$ stands for $\rho\rightarrow 0$) in the respective cases of ultrashort LG and X vortices. Thus for $|l|>0$, $\bar\omega_E(\epsilon)>\bar\omega(0)=\bar\omega_a$ for general $\hat a_\omega$, i. e., there is always a blue shift towards the vortex center, and this blue shift is more pronounced for X vortices than for LG vortices. Small blue shifts of the carrier oscillations have been previously described, for example, for pulsed Gaussian beams about the beam axis at the far field \cite{SHEPPARD,KAPLAN,PORRAS5}. Here on-axis blue shift is permanent along the propagation direction and is monotonously enhanced as the magnitude of the topological charge $|l|$ increases, as illustrated in Figs. \ref{Fig5}(c) and (d) for the same three particular spectra of single-cycle pulses. For spectra without a cut-off frequency, as the power-exponential and Gaussian spectra, the mean frequency about the vortex grows without bound with the magnitude of the topological charge, and for spectra with a cut-off frequency $\omega_c$, grows up to it. These huge blue shifts should be observable in experiments, e. g., with the spectral bandwidth of a single-cycle pulse in the near infrared and topological charge $|l|=20$, the mean frequency in the bright caustic is indeed in the near infrared, but is in the visible region or in the ultraviolet close to the vortex of ultrashort LG and X vortices.

Also in \cite{CONTI1}, a slow shortening of the pulse with increasing $|l|$ close to the vortex of ultrashort X vortices with power-exponential spectra is reported, which together with the more pronounced decrease of the mean period, $2\pi/\bar \omega_E(\epsilon)$, results in an increase of the number of oscillations with as $|l|$ grows. This effect is seen here to be almost identical for ultrashort LG vortices with power-exponential spectra. However, it also follows also from our analysis that pulse shortening or lengthening is not a general temporal-OAM coupling effect, but is dictated by the particular spectrum $\hat a_\omega$, and that the only two general temporal-OAM coupling effects close to the vortex are blue shift and increase of the number of oscillations with $|l|$. In a sense, ultrafast LG and X vortices always approach a monochromatic behavior in the vicinity of the vortex as their topological charge increases.

Indeed, the definition of bandwidth in Eq. (\ref{GEBW}) for the spectra $\hat E_\omega(\rho)$ of ultrafast LG and X vortices close to the center leads to the expressions of the bandwidth
\begin{eqnarray}
\Delta \omega_E(\epsilon) &=& 2\sqrt{\bar\omega(|l|)[\bar\omega(|l|+1)-\bar\omega(|l|)]}\,, \\
\Delta \omega_E(\epsilon) &=& 2\sqrt{\bar\omega(2|l|)[\bar\omega(2|l|+1)-\bar\omega(2|l|)]}
\end{eqnarray}
for ultrafast LG and X vortices, respectively. As in the examples in Fig. \ref{Fig6}(a) with the same three spectral shapes $\hat a_\omega$ of single-cycle pulses as in Fig. \ref{Fig5}, the bandwidth close to the vortex may increase with $|l|$, decrease, or reach a constant value, the corresponding pulse durations then behaving in reverse, as illustrated in Figs. \ref{Fig6}(c), (d) and (e) for the three type of spectral shapes. However, the relative bandwidths,
\begin{eqnarray}
\frac{\Delta \omega_E(\epsilon)}{\bar\omega_E(\epsilon)} &=& 2\sqrt{\frac{\bar\omega(|l|+1)}{\bar\omega(|l|)}-1}\,, \\
\frac{\Delta \omega_E(\epsilon)}{\bar\omega_E(\epsilon)} &=& 2\sqrt{\frac{\bar\omega(2|l|+1)}{\bar\omega(2|l|)}-1}
\end{eqnarray}
for ultrafast LG and X vortices, respectively, always go to zero with growing $|l|$, as in Fig. \ref{Fig6}(b), since $\bar\omega(m+1)/\bar\omega(m)$ approaches unity with increasing $m$. Hence, the number $N$ of mean or carrier periods, $2\pi/\bar\omega_E(\epsilon)$, in the full pulse duration, $2\Delta t_E(\epsilon)$, with $\Delta t_E(\epsilon)\ge 2/\Delta \omega_E(\epsilon)$, or $N= \Delta t_E(\epsilon)\bar\omega_E(\epsilon)/\pi$, always satisfies $N \ge (2/\pi)\bar\omega_E(\epsilon)/ \Delta\omega_E(\epsilon)$, with the right hand side of the inequality growing without bound with increasing $|l|$, regardless the choice of $\hat a_\omega$. This is why the pulse shapes close two the vortex in Figs. \ref{Fig6}(c), (d) and (e) always contains more oscillations than $a(t')$, irrespective that the pulse is longer or shorter.

\section{Conclusions}

In this manuscript we have reported a unified treatment of ultrashort vortices with propagation-invariant temporal shape for studying the universal and diverse forms of their coupling between the temporal and orbital angular momentum degrees of freedom, a kind of spatiotemporal coupling peculiar to ultrashort vortices. This study completes previous studies and clarifies their seemingly contradictory results.

We have shown that orbital angular momentum introduces strong correlation between the amount of spatial twisting, the local frequency, and the pulse duration. The coupling manifests itself in two ways in different regions of the three-dimensional energy distribution. In the high-intensity caustic surface surrounding the vortex, the $l$ dependent lower bound to the pulse duration previously reported for ultrashort LG vortices holds similarly for ultrashort X vortices. As a consequence, the pulse duration of both ultrashort LG and X vortices must increase with the angular momentum.
This fact may have an impact in multilevel-OAM pulsed transmission systems. In the low-intensity region, in the proximity of the vortex phase singularity, we found that a blue shift with an increase of the number of oscillations with respect to those at the bright caustic surface occurs not only for ultrashort X vortices, as previously reported, but also for ultrashort LG vortices. Pulse shortening or lengthening in the vicinity of the vortex is not a general feature of the temporal-OAM coupling, but depends on the particular model.
Higher frequencies are then located in low-energy-density regions close to the vortex. The local blue shift increases with the topological charge and is larger for ultrafast X vortices than for LG vortices.

It is an open question if this remarkable space-time correlation in the dark regions has fundamental implications in quantum optics at the single-photon level, or for entangled beams. We think that the blue shift can be measured with synthesized propagation-invariant pulses, ranging from terahertz to visible. We also believe that the frequency shift may be evident in high field phenomena, as in high harmonic generation, where the amount of angular momentum grows with the harmonic order, and where --so far-- spatiotemporal coupling effects and the internal structure of the generated beams have been overlooked.

\end{document}